\documentclass[letterpaper, 10 pt, conference]{ieeeconf}  % Comment this line out if you need a4paper

\IEEEoverridecommandlockouts                              % This command is only needed if
\usepackage{amsfonts}
\usepackage{mathrsfs}
\usepackage{amssymb}
\usepackage{extarrows}
\usepackage{color}
\usepackage{cite}

\def\rep#1{(\ref{#1})}

\newcommand{\R}{\mathbb{R}}

\def\send#1#2{\stackrel{#1}{\hbox to #2{\rightarrowfill}}}
\def\-{\!\!\!\!\!-}

 \def\qed{ \rule{.1in}{.1in}}
\def\eq#1{\begin{equation}#1\end{equation}}

\def\scr#1{{\cal #1}}

\newcommand{\matt}[1]{\begin{bmatrix}#1\end{bmatrix}}
\newcommand{\dfb}{\stackrel{\Delta}{=}}

\newtheorem{lemma}{Lemma}

\newtheorem{proposition}{Proposition}

\def\qed{ \rule{.1in}{.1in}}

\def\R{{\rm I\!R}}

\newcounter{seqn}[equation]
\def\theseqn{\arabic{equation}\alph{seqn}}

\def\endseqn{\eqno \@seqnnum
$$\ignorespaces}
\def\@seqnnum{(\theseqn)}

\newskip\mcentering \mcentering=0pt plus 1000pt minus 1000pt

\def\meqalignno#1{
\halign to\displaywidth{
    \hbox to 0pt{\kern\displaywidth\llap{$##$}\hss}\tabskip=\mcentering
    &\hfil$\displaystyle{##}$\tabskip=\mcentering
   &&$\displaystyle{{}##}$\hfil\tabskip=\mcentering
    \crcr
    #1\crcr}}

\def\rep#1{(\ref{#1})}

\def\eq#1{\begin{equation}#1\end{equation}}
\def\dspace{\multiply\normalbaselineskip 150
		  \divide\normalbaselineskip 100 \normalbaselines
		  \csname @@normalbaselineskip\endcsname\normalbaselineskip}
\def\sspace{\multiply\normalbaselineskip 200
		 \divide\normalbaselineskip 300 \normalbaselines
		 \csname @@normalbaselineskip\endcsname\normalbaselineskip}
\def\sdspace{\multiply\normalbaselineskip 160
		 \divide\normalbaselineskip 150 \normalbaselines
		 \csname @@normalbaselineskip\endcsname\normalbaselineskip}

\def\@{\tilde}

\def\3dot#1{\buildrel\textstyle...\over#1}

\title{\LARGE \bf
A Distributed Observer for a Discrete-Time Linear System }

\author{Lili Wang, Ji Liu, A. Stephen Morse, and Brian D. O. Anderson
\thanks{This work was  supported by  NSF grant  1607101.00,
  AFOSR grant  FA9550-16-1-0290, ARO grant  W911NF-17-1-0499, and Australian Research Council Grant DP-160104500.}% <-this % stops a space
\thanks{L. Wang and A.S. Morse  are with the Department
of Electrical Engineering, Yale University ({\tt\small \{lili.wang,
as.morse\}@yale.edu}). J.~Liu is with the Department of Electrical
and Computer Engineering, Stony Brook University ({\tt\small
ji.liu@stonybrook.edu}). B.D.O. Anderson is with the Research School
of Engineering, Energy and Materials Engineering, Australian
National University, Canberra, Australia; also with Hangzhou Dianzi
University, China; and with CSIRO Data-61. ({\tt\small
brian.anderson@anu.edu.au}).
}%
}

\begin{document}
\maketitle \thispagestyle{empty}

\begin{abstract}

A simply structured distributed  observer is described for
estimating the state
 of a discrete-time, jointly observable, input-free, linear system
 whose sensed outputs are distributed across a
 time-varying network. It is explained how to construct the local estimators which comprise the observer so that
  their state estimation errors all converge exponentially fast to zero at a fixed, but arbitrarily chosen rate provided
   the network's graph is strongly connected for all time.
This is accomplished by  exploiting several well-known properties of
invariant
  subspaces plus several kinds of  suitably defined matrix norms.

\end{abstract}

%\cite{Horn} \cite{graph}

\section{Introduction}\label{intro}

With  the  growing interest in  sensor networks and multi-agent
systems, the problem of estimating the state of a dynamical system
whose measured outputs are distributed across a network has been
under  study in one form or another
 for a number of years
  \cite{KhanAli2011ACC,shamma,carli,xxx,saber2,bullo.observe,sanfelice}.
    Despite this, only quite recently have provably correct distributed state estimators begun to
 emerge which solve this problem under reasonably non-restrictive assumptions
\cite{martins,TAC.17,MitraPurdue2016,
Kim2016CDC,trent,trent2,CDC17.1,ACC19}.

In its simplest form, the discrete-time version of the distributed
state estimation
 problem
 starts with
a network of $m>1$ agents labeled $1,2,\ldots,m$ which are able to
receive information from their neighbors. Neighbor relations are
characterized by a directed graph $\mathbb{N}$, which may or may not
depend on time,  whose
     vertices correspond to agents and whose arcs depict neighbor relations.
Each agent $i$  senses a
   signal $y_i\in\R^{s_i},\;i\in\mathbf{m} = \{1,2,\ldots,m\}$ generated by a discrete-time system of the form
$x(\tau+1)=Ax(\tau),\;y_i(\tau) = C_ix(\tau),\;i\in \mathbf{m}$ and
$x\in\R^n$. It is typically  assumed that $\mathbb{N}$ is strongly
connected and that the system  is jointly observable. It is
invariably assumed that each agent  receives certain real-time
signals from its
 neighbors although what is received can vary from one problem formulation to the next.
 In all formulations, the goal is to devise local estimators, one for each agent, whose outputs
 are all asymptotically correct estimates of $x$.
 The local estimator dynamics for agent $i$ is typically assumed to depend only on the pair $(C_i,A)$
 and certain properties of $\mathbb{N}$.
The problem is basically the same in continuous time, except that
rather than the discrete-time model just described,  the
 continuous-time model
$\dot{x}=Ax,\;y_i(t) = C_ix, \; i\in \mathbf{m}$  is considered
instead.

One way to try to address the estimation problem is to recast it as
a discrete-time
 classical decentralized control
problem \cite{corfmat} as was done in \cite{martins}. Following this
approach, it is possible to devise  a provable correct procedure for
crafting a distributed linear filter with a prescribed spectrum
which solves
 the continuous-time version of the problem  assuming
  $\mathbb{N}$ is a constant strongly connected graph \cite{TAC.17}; the same procedure is easily modified to
  deal with the discrete-time version of the problem.
   Prompted by work in \cite{Kim2016CDC}, an entirely different and simpler approach
    to the continuous-time version of the estimation problem was developed in
   \cite{trent}. The same approach was simplified still further in \cite{ACC19} by exploiting certain
   well-known properties of
    invariant subspaces. There are however two distinct limitations of the types of estimators
     discussed  in \cite{Kim2016CDC,trent,ACC19}. First, as they stand these estimators cannot deal with
      time-varying neighbor graphs. Second,  there does not appear to be a way to easily modify these estimators to
     address the discrete-time state estimation problem; this is because the continuous-time estimators rely on a
     ``high gain'' concept for which there is no discrete-time counterpart.
      Despite these limitations, there is a very useful idea in these papers, stemming from the work in
      \cite{Kim2016CDC}, which can be used to advantage in developing a discrete-time solution to the problem.
Roughly speaking, the idea is to using the invariance of the
unobservable spaces of the  the pairs $(C_i,A)$ to ``split'' the
estimators into two parts - one for which conventional spectrum
assignment tools can be used to control convergence rate and the
other for which convergence rate can be controlled by switching and
averaging.

This paper is organized as follows. Certain basic properties of
invariant subspaces are reviewed in
 \S\ref{in}. The specific problem to be addressed is then  formulated in \S\ref{form}. In \S\ref{obs}
  the observer which solves this problem is described. The error
  model needed to analyze the observer is developed in \S\ref{errors}. Finally in \S\ref{switch},
  several techniques are outlined for picking the number of switches required between ``event times''
   in order to achieve  a prescribed convergence rate.

\subsection{Invariant Subspaces}\label{in}
Throughout this paper certain basic and well-known algebraic
properties of invariant subspaces will be exploited. To understand
what they are,
  let $A$ be any square matrix, and suppose $\scr{V}$ is an $A$-invariant subspace. Let $Q$
  be any full row rank matrix whose kernel is $\scr{V}$ and  suppose that  $V$ is any ``basis matrix''
 for $\scr{V}$; i.e., a matrix whose columns form a basis for $\scr{V}$.
Then the linear equations $$QA =\bar{A}_{V}Q\;\;\;\;\;
\text{and}\;\;\;\;\; AV=VA_{V}$$
   have unique solutions $\bar{A}_V$ and $A_V$ respectively. Let $V^{-1}$ be any left inverse of $V$ and let
    $Q^{-1}$ be that right inverse of $Q$ for which $V^{-1}Q^{-1} =0 $.
    Then
 %   $\matt{Q^{-1} & V}$ is the inverse of $\matt{Q\\V^{-1}}$,  $$V^{-1}A = \widehat{A}_VQ +
  %  A_{V}V^{-1},$$
   % and
$$A = H^{-1}\matt{\bar{A}_V & 0\\ \widehat{A}_V & A_{V}}H$$
where
$$H = \matt{Q \\ \\ V^{-1}}$$
and
 $\widehat{A}_V = V^{-1}AQ^{-1}$.
Use will be made of these simple algebraic facts in the sequel.

%\subsection{Background}

\section{Problem}\label{form}

 We are interested in a  time-varying network
of $m>1$ agents labeled $1,2,\ldots,m$ which are able to receive
information from their neighbors where by a {\em neighbor} of agent
$i$ is meant any agent in agent $i$'s reception range.
 We write $\scr{N}_i(t)$ for the set of labels of agent $i$'s neighbors at real time $t$ and
 take agent $i$
 to be a neighbor of itself for all $t$. Relations between neighbors are characterized by a directed graph
  $\mathbb{N}(t)$ with $m$ vertices and a set of arcs defined so that there
  is an arc from vertex $j$ to vertex $i$ whenever
   agent $j$ is a neighbor of agent $i$.
Each agent $i$ can sense a discrete-time
   signal $y_i(\tau)\in\R^{s_i}$ at {\em event times} $\tau T$, $\tau = 0, 1,2,\ldots $
     where $T$ is a positive constant;
 for  $i\in\mathbf{m} \dfb \{1,2,\ldots,m\}$ and $\tau=0,1,2,\ldots $
\eq{y_i(\tau)   =  C_ix(\tau) ,\;\;\;\;\;x(\tau +1) =Ax(\tau )
\label{syss}} and $x\in\R^n$. We assume throughout that
$\mathbb{N}(t)$ is strongly connected and that the system defined by
\rep{syss} is {\em jointly observable}; i.e., with $C =
\begin{bmatrix}C_1' &C_2' & \cdots & C_m'\end{bmatrix}'$, the matrix
pair $(C,A)$ is observable. Joint observability is equivalent to the
requirement that
$$\bigcap_{i\in\mathbf{m}}\scr{V}_i = 0$$%\label{ob}}
where $\scr{V}_i$ is the {\em unobservable space} of $(C_i,A)$; i.e.
$\scr{V}_i = \ker \begin{bmatrix}C_i' &(C_iA)' & \cdots
&(C_iA^{n-1})'\end{bmatrix}'$. As is well known, $\scr{V}_i$ is
 the largest $A$-invariant subspace contained in the
kernel of $C_i$.

Each agent $i$  is to  estimate $x$ using a dynamical system whose
output  $x_i(\tau)\in\R^n$
 is to be an asymptotically correct estimate of $x(\tau)$ in the sense that the estimation error
$x_i(\tau)-x(\tau)$ converges to zero  as $\tau\rightarrow \infty$
as fast as $\lambda^{\tau} $ does,
 where $\lambda$ is an
arbitrarily chosen but fixed
 positive number\footnote{For the type of observer to be developed, finite-time convergence is not possible.} less than $1$. To accomplish this it is assumed that
 the information
agent $i$ can receive from neighbor $j$ at event time  $\tau T$ is
$x_j(\tau)$. It is further
 assumed that agent $i$
can also receive certain additional information from its neighbors
at a finite number of
 times  between each successive pair of  event times; what this  information is  will be specified below.

\section{The Observer}\label{obs} In this paper it will be assumed that each agent's neighbors do not
 change between event times. In other words,  for $i\in\mathbf{m}$,  $$\scr{N}_i(t) =
  \scr{N}_i(\tau T),\;\;\;\; t\in [\tau T,(\tau+1)T),\;\;\;\;\tau = 0,1,2,\ldots $$
With this assumption, the observer to be considered
 consists of $m$ private estimators, one for each agent.
  The estimator  for agent $i$ is  of the form
\eq{ x_i(\tau+1) =
 (A+K_iC_i)\bar{x}_i(\tau) -K_iy_i(\tau)\label{est1}}
where $\bar{x}_i(\tau)$ is an ``averaged state'' computed
recursively  during the real time interval $[\tau T,\; (\tau+1)T)$
using the update  equations
\begin{eqnarray}
z_i(0,\tau) &=&x_i(\tau)\label{p1}\\
z_i(k,\tau) &=& (I-P_i)z_i(k-1,\tau)\nonumber\\ &+& \frac{1}{m_i(\tau)}P_i\sum_{j\in\scr{N}_i(\tau T)}z_j(k-1,\tau),\;\;k\in \mathbf{q}\\
\bar{x}_i(t) &= &z_i(q,\tau)\label{p3}\end{eqnarray} Here
$m_i(\tau)$ is the number of labels in $\scr{N}_i(\tau T)$, $q$ is a
suitably defined positive integer,
 $\mathbf{q}\dfb\{1,2,\ldots, q\}$,  and   $P_i$ is the orthogonal projection on the unobservable space of $(C_i,A)$.
Each matrix  $K_i$ is defined as follows.

    For  fixed $i\in\mathbf{m}$,  write $Q_i$ for any full  rank matrix whose kernel
    is the unobservable space of $(C_i,A)$, and let $\bar{C}_i$ and $\bar{A}_i$ be the unique solutions to
    $\bar{C}_iQ_i = C_i$ and $Q_iA=\bar{A}_iQ_i$  respectively.  Then  the matrix pair $(\bar{C}_i,\bar{A}_i)$ is
     observable.  Thus by using a standard spectrum assignment algorithm,
     a matrix $\bar{K}_i$ can be chosen to ensure that  the convergence
     of $(\bar{A}_i + \bar{K}_i\bar{C}_i)^{\tau}$ to zero as $\tau\rightarrow \infty$
     is as fast as the convergence to zero
      of  $\lambda ^{\tau}$ is.
      Having chosen such $\bar{K}_i$, $K_i$
      is then defined to be $K_i = Q_i^{-1}\bar{K}_i$ where $Q_i^{-1}$ is a right inverse for $Q_i$.
The definition implies that $Q_i(A+K_iC_i) =
(\bar{A}_i+\bar{K}_i\bar{C}_i)Q_i$ and that
$(A+K_iC_i)\scr{V}_i\subset \scr{V}_i$. The latter, in turn, implies
that there is a unique matrix  $A_i$ which satisfies $(A+K_iC_i)V_i
= V_iA_i$ where $V_i$ is a basis matrix\footnote{For simplicity, we
       assume that the columns of $V_i$ constitute an orthonormal basic for $\scr{V}_i$
        in which case  $P_i = V_iV_i'$.} for $\scr{V}_i$.
% Note in addition that
        %$\sigma(A+K_iC_i) = \sigma(\bar{A}_i +\bar{K}_i\bar{C}_i)\cup\sigma{A_i}$.
        To explain what needs to
         be considered in choosing
        $q$ it is necessary to describe  the structure of the  ``error model'' of  the overall observer. This will be done next.

\section{The Error Model}\label{errors}

For $i\in\mathbf{m}$, write $e_i(\tau)$  for the {\em state
estimation error} $e_i(\tau) = x_i(\tau)-x(\tau )$. In view of
\rep{est1},
$$e_i(\tau+1) = (A+K_iC_i)\bar{e}_i(\tau)$$
where $\bar{e}_i(\tau) =\bar{x}_i(\tau)-x(\tau)$.  Moreover if  %$\epsilon_i(0,t) =e_i(t)$ and
 $\epsilon_i(k,\tau) \dfb z_i(k,\tau)-x(\tau),\;k\in\{0,1,\ldots, q\}$
then
\begin{eqnarray}
\epsilon_i(0,\tau) & =& e_i(\tau)\\
\epsilon_i(k,\tau) &=& (I-P_i)\epsilon_i(k-1,\tau)\nonumber\\
& +&\frac{1}{m_i(\tau)}P_i\sum_{j\in\scr{N}_i(\tau T)}\epsilon_j(k-1,\tau),\;\;k\in \mathbf{q}\\
\bar{e}_i(\tau) &= &\epsilon_i(q,\tau)\end{eqnarray} because of
\rep{p1} -- \rep{p3}. It is possible to combine   these $m$
subsystems  into a single system. For this let
  $e = $ column $\{e_1,e_2,\ldots, e_m\}$, define
$\bar{A} = $ block diagonal $\{A+K_1C_1
,A+K_2C_2,\ldots,A+K_mC_m\}$,
 $P = $ block diagonal $\{P_1 ,P_2,\ldots, P_m\}$ and write $S(\tau)$ for the stochastic matrix
$S(\tau) = D_{\mathbb{N}(\tau T)}^{-1}A'_{\mathbb{N}(\tau T)}$ where
$A_{\mathbb{N}(\tau T)}$ is
 the adjacency matrix of $\mathbb{N}(\tau T)$ and  $ D_{\mathbb{N}(\tau T)}$
 is the diagonal matrix whose $i$th diagonal entry is the in-degree
  of $\mathbb{N}(\tau T)$'s $i$th vertex. Note that $\mathbb{N}(\tau T )$ is the graph\footnote{The {\em graph} of an $n\times n$ matrix $M$ is  that directed graph
  on $n$ vertices possessing
 a directed arc from vertex  $i$ to vertex $j$ if $m_{ij}\neq 0$ \{p. 357, \cite{horn1}.\}}
 of $S'(\tau)$ and that the diagonal entries of $S'(\tau)$
 are all positive because each agent is a neighbor of itself.

Let $\bar{e}(\tau) = $ column
$\{\bar{e}_1(\tau),\bar{e}_2(\tau),\ldots, \bar{e}_m(\tau)\}$ and
$\epsilon(k,\tau) =
\text{column}\{\epsilon_1(k,\tau),\epsilon_2(k,\tau),\ldots
,\epsilon_m(k,\tau)\}$. Then
$$e(\tau +1) = \bar{A}\bar{e}(\tau)$$
 and
\begin{eqnarray*}
\epsilon(0,\tau) &=&e(\tau)\\
\epsilon(k,\tau) &=& (I_{mn}-P(I_{mn}-\bar{S}(\tau)))\epsilon(k-1,\tau),\;k\in\mathbf{q}\\
\bar{e}(\tau) &= &\epsilon(q,\tau)\end{eqnarray*} where
$\bar{S}(\tau ) = S(\tau)\otimes I_n$; here $\otimes $ denotes
Kronecker product, and   $I_n$ and $I_{mn}$ are the  $n\times n$ and
$mn\times mn$  identity matrices respectively. Clearly
$$\bar{e}(\tau) = (I_{mn}-P(I_{mn}-\bar{S}(\tau)))^qe(\tau)   $$
so \eq{e(\tau+1) = \bar{A}(I_{mn}-P(I_{mn}-\bar{S}(\tau)))^qe(\tau)
\label{error}}

Our aim is to explain why  for $q$ sufficiently large, the
time-varying matrix
 $\bar{A}(I_{mn}-P(I_{mn}-\bar{S}(\tau)))^q$ appearing in \rep{error} is a discrete-time stability matrix
for which the product \eq{\Phi(\tau) = \prod_{s =
1}^{\tau}\bar{A}(I_{mn}-P(I_{mn}-\bar{S}(s)))^q\label{phi}}
converges to zero as $\tau\rightarrow\infty $ as fast as $\lambda
^{\tau}$ does. As a first step towards this end, note that  the
subspace $\scr{V} = \scr{V}_1\oplus\scr{V}_2\oplus \cdots \oplus
\scr{V}_m$
 is
$\bar{A}$ - invariant  because $(A+K_iC_i)\scr{V}_i\subset
\scr{V}_i,\;i\in\mathbf{m}$. Next, let $Q = $ block diagonal $
\{Q_1, Q_2,\ldots ,Q_m\}$ and $V = $ block diagonal $ \{V_1,
V_2,\ldots ,V_m\}$ in which case $Q$ is a full rank matrix whose
kernel is $\scr{V}$  and $V$ is a basis matrix for $\scr{V}$ whose
 columns form an orthonormal set.
 It follows that $P = VV'$, and that
\begin{eqnarray}
Q\bar{A}& = &\bar{A}_VQ\label{prr1}\\
 \bar{A}V & = & V\tilde{A}\label{prr2}
 \end{eqnarray}
  where
\eq{\bar{A}_V =  \text{block diagonal}\;
\{\bar{A}_1+\bar{K}_1\bar{C}_1 ,\ldots,
\bar{A}_m+\bar{K}_m\bar{C}_m\}\label{sunday}} and
$$ \tilde{A} =\text{block diagonal}\;\{A_1,A_2,\ldots ,A_m\};$$
as before, $(A+K_iC_i)V_i=V_iA_i$. Moreover
\begin{eqnarray}Q(I_{mn}-P(I_{mn}-\bar{S}(\tau)))^q &=& Q\label{pr1}\\
(I_{mn}-P(I_{mn}-\bar{S}(\tau)))^qV& =& V(
V'\bar{S}(\tau)V)^q\label{pr2}\end{eqnarray} Note that \rep{pr1}
holds  because $QP = 0$.  To understand why \rep{pr2} is true, note
first that $(I_{mn}-P(I_{mn}-\bar{S}(\tau)))V =  V(I_{\bar{n}} -
V'(I_{mn} - \bar{S}(\tau))V)$ because $P =VV'$; here $\bar{n} =
\dim(\scr{V})$. But $I_{\bar{n}} - V'(I_{mn} - \bar{S}(\tau))V =
V'\bar{S}(\tau)V$ because
  $V'V = I_{\bar{n}}$. Thus \rep{pr2} holds for $q=1$; it follows by induction that \rep{pr2}
  holds for any positive integer $q$.

Using \rep{prr1} -- \rep{pr2}, one obtains the equations
\begin{eqnarray}Q\bar{A}(I_{mn}-P(I_{mn}-\bar{S}(\tau)))^q &=& \bar{A}_VQ\label{xpr1}\\
\bar{A}(I_{mn}-P(I_{mn}-\bar{S}(\tau)))^qV& =& VA_V(\tau)
\label{xpr2}\end{eqnarray} where \eq{A_V(\tau) =
\tilde{A}(V'\bar{S}(\tau)V)^q\label{ench}} These equations imply
that
\eq{\bar{A}(I_{mn}-P(I_{mn}-\bar{S}(\tau)))^q = H^{-1}\matt{\bar{A}_V & 0\\
\hat{A}_V(\tau) & A_V(\tau)}H\label{as}} where $$H =\matt{Q \\ \\
V^{-1}}$$ and $\widehat{A}_V(\tau) =
V^{-1}\bar{A}(I_{mn}-P(I_{mn}-\bar{S}(\tau)))^qQ^{-1}$.

Since the spectrum of each
$\bar{A}_i+\bar{K}_i\bar{C}_i,\;i\in\mathbf{m}$, is assignable with
$\bar{K}_i$, and $\widehat{A}_V(\tau)$  is a bounded matrix, to show
that  for suitably defined $\bar{K}_i$ and  $q$ sufficiently large,
the matrix $\Phi(\tau )$ defined in \rep{phi}
 converges to zero as fast as $\lambda^{\tau}$ does, it is sufficient to show that for $q$ sufficiently large,
 $A_{V}(\tau )$ is a discrete-time stability matrix whose state-transition matrix converges to zero
as fast as $\lambda^{\tau}$ does. To accomplish this, use will be
made of the following results.

%\begin{proposition}
%For each fixed value of $\tau$, $V'\bar{S}(\tau)V$ is a discrete-time stability matrix
%and
%\eq{(V'\bar{S}(\tau)V)' R(\tau)(V'\bar{S}(\tau)V) - R(\tau) < 0\label{ly}}\label{mp} \end{proposition}
%where $R(\tau) =  \Pi(\tau} \otimes I_n$.

\begin{lemma} Let $M$ be an $m\times m$ row stochastic matrix  whose transpose has a strongly connected graph.
There exists  a diagonal matrix $\Pi_M$ whose diagonal entries are
positive for which the matrix $L_M= \Pi_M- M'\Pi_M M $ is positive
semi-definite;  moreover $L_M\mathbf{1} = 0$ where
  $\mathbf{1}$ is the $m$-vector of $1$s. If, in addition,
 the diagonal entries of $M$ are all positive, then
 the  kernel of $L_M$ is one-dimensional.
 \label{brian} \end{lemma}

\noindent{\bf Proof of Lemma \ref{brian}:} Since $M$ is a stochastic
matrix, it must have
 a spectral  radius of $1$ and an  eigenvalue  at $1$ as must $M'$.  Moreover,  since
 the graph of $M'$ is strongly
 connected, $M'$ is
  irreducible \{Theorem 6.2.24, \cite{horn1}\}. Thus by the Perron-Frobenius Theorem there
   must be a positive vector $\pi$
such that $M'\pi= \pi$. Without loss of generality, assume
  $\pi$  is normalized so that the sum of its entries equals $1$; i.e., $\pi$ is a probability vector.
   Let $\Pi_M $ be that  diagonal matrix
  whose diagonal entries are the entries of $\pi$. Then $\Pi_M\mathbf{1} = \pi$.

 Since  $M\mathbf{1} = \mathbf{1}$,  $\Pi_M \mathbf{1} = \pi$, and
$M'\pi = \pi$,   it must be true that $M'\Pi_M M\mathbf{1} = \pi$
and thus that $L_M\mathbf{1} = 0$. To show that $L_M$ is
positive-semidefinite note first that
 $L_M$ can also be written as $L_M=D-\hat{A}$ where $D$ is a
  diagonal matrix whose diagonal entries are the diagonal
   entries of $L_M$
  and $\hat{A}$ is the nonnegative matrix $\hat{A} = D- L_M$.
   As such, $L_M$ is the generalized Laplacian  \cite{graph}
   of that
  simple undirected graph $\mathbb{G}$  whose adjacency matrix is the matrix which results when the
  nonzero entries $a_{ij}$
  in $\hat{A}$ are replaced by ones. Since  $L_M$ can also be written as
   $$L_M=\sum_{(i,j)\in\scr{E}} a_{ij}(e_i-e_j)(e_i-e_j)'$$ where $e_i$ is the $i$th unit vector
    and $\scr{E}$ is the edge
   set of $\mathbb{G}$, $L_M$ is
    positive semi-definite as claimed.

Now suppose that the diagonal entries of $M$ are all positive. Then
the diagonal entries  of $M'\Pi_M $ must also all be positive. It
follows that
 every arc in the graph of $M'$ must be an arc in the graph of $M'\Pi_M M$ so the graph of $M'\Pi_M M$ must
  be strongly connected. Since $I-\Pi_M$ is a nonnegative matrix, the graph of $M'\Pi_M M$ must be
 a spanning subgraph
   of the graph of $I-\Pi_M +M'\Pi_M M$.
 Since $I-L_M = I-\Pi_M +M'\Pi_M M$ and the graph of $M'\Pi M$ is strongly connected, the graph of $I-L_M$ must
  be strongly connected as well.
But $I-L_M$ is a nonnegative matrix so it must be irreducible. In
addition, since
   $(I-L_M)\mathbf{1} = \mathbf{1}$, the row sums of $(I-L_M)$ all equal one. Therefore the infinity norm of
    $I-L_M$ is one so its spectral radius is no greater than $1$. Moreover $1$ is an eigenvalue of $I-L_M$.
     Thus
by the  Perron-Frobenius Theorem, the geometric multiplicity of this
eigenvalue is one. It follows that the geometric multiplicity of the
eigenvalue of $L_M$ at $0$ is also one;
 ie, the dimension of the kernel of $L_M$ is one as claimed. \qed

\begin{proposition}
For each fixed value of $\tau$, \eq{(V'\bar{S}(\tau)V)'
R(\tau)(V'\bar{S}(\tau)V) - R(\tau) < 0\label{ly}} where $R(\tau )$
is the positive definite matrix, $R(\tau) =  V'(\Pi_{S(\tau)}
\otimes I_n)V$.

\label{mp} \end{proposition}

Note that \rep{ly} shows that for each fixed $\tau$, $x'R(\tau)x$  a
discrete-time Lyapunov function for the equation $w(k+1)=V'\bar
S(\tau)Vw(k)$.  Thus for fixed $\tau$,
  $V'\bar{S}(\tau)V$ is a discrete-time stability matrix.

\noindent{\bf Proof of Proposition \ref{mp}:} Fix $\tau $ and write
$S$ for $S(\tau)$ and $\bar{S}$ for $\bar{S}(\tau)$. Note that the
graph of $S'$, namely $\mathbb{N}$,   is strongly connected.
 In view of Lemma \ref{brian}, the matrix $L = \Pi_S-S'\Pi_S S$
 is positive semi-definite and $L\mathbf{1} = 0$. Moreover, since
the diagonal entries of $S$ and thus $S'$  are all positive, the
kernel of $L$ is one-dimensional.

Write $R$ for $R(\tau )$.    %$R =  V'\bar{\Pi} V $ where  $\bar{\Pi} = \Pi(\tau) \otimes I_n$.
To prove the proposition  it is enough to show that the matrix \eq{
Q = R- (V'\bar{S}'V)R(V'\bar{S}V)\label{1}} is positive definite.

To proceed, set   $\bar{L} = L\otimes I_n$ in
 which case $\bar{L}$ is positive semi-definite because $L$ is. Moreover,
 $\bar{L} = \bar{\Pi}- \bar{S}'\bar{\Pi }\bar{S}$ where $\bar{\Pi} = \Pi_{S}\otimes I_n$.
Note that  that $VRV'  = P\bar{\Pi}P$ where $P$ is the orthogonal
projection matrix $P=VV'$. Clearly $VRV' =
P\bar{\Pi}^{\frac{1}{2}}\bar{\Pi}^{\frac{1}{2}}P$. Note that both
$P$ and $\bar{\Pi}^{\frac{1}{2}}$ are block diagonal matrices with
corresponding diagonal blocks of the same size. Because of this and
the
 fact that each diagonal block in $\bar{\Pi}^{\frac{1}{2}}$
 is a scalar times and identity matrix, it must be true that $P$ and $\bar{\Pi}^{\frac{1}{2}}$ commute; thus
$P\bar{\Pi}^{\frac{1}{2}} = \bar{\Pi}^{\frac{1}{2}}P$. From this and
the fact that $P$ is
 idempotent, it follows that
$VRV' =\bar{\Pi}^{\frac{1}{2}}P\bar{\Pi}^{\frac{1}{2}}$.
 Clearly $\bar{\Pi}^{\frac{1}{2}}P\bar{\Pi}^{\frac{1}{2}}\leq
\bar{\Pi}^{\frac{1}{2}}\bar{\Pi}^{\frac{1}{2}}$ so $VRV' \leq
\bar{\Pi}$. It follows using \eqref{1} that $Q\geq
R-V'\bar{S}'\bar{\Pi}\bar{S}V = R+ V'\bar{L}V - V'\bar{\Pi}V $.
Therefore \eq{Q\geq V'\bar{L}V\label{water}} In view of this, to
complete the proof it is enough to show that $V'\bar{L}V$ is
positive definite.

Since $\bar{L}$ is positive semi-definite, so is $V'\bar{L}V$. To
show that $V'\bar{L}V$ is positive definite, let $z = \text{column}
\{z_1,z_2,\ldots, z_m\}$  be any vector such that $z'V'\bar{L}Vz=0$.
Then $\bar{L}Vz = 0$. Since the kernel  of  $L$ is spanned
$\mathbf{1}$, the kernel of $\bar{L}$   must  be spanned by
$\mathbf{1}\otimes I_n$. It follows that
$V_iz_i=V_jz_j,\;i,j\in\mathbf{m}$.  But because of joint
observability, $\bigcap_{i\in\mathbf{m}} \scr{V}_i = 0$ so $V_iz_i =
0,\;i\in\mathbf{m}$. Thus $z_i = 0,\;i\in\mathbf{m}$
 so $z=0$.
Therefore   $V'\bar{L}V$ is positive definite. Therefore $Q$ is
positive definite because of  \rep{water}. From this and \rep{1}
 it follows that \rep{ly} is true.
\qed

\section{Choosing $q$}\label{switch}

In what follows it will be assumed that each $\bar{K}_i$ has been
selected so that  the the matrix
 $\bar{A}_V$ defined by \rep{sunday}, is such that $\bar{A}_V^{\tau}$ converges
  to zero  as $\tau\rightarrow \infty$ as fast as $\lambda^{\tau}$ does.
 This can be done using standard
spectrum assignment techniques to  make the spectral radius of
$\bar{A}_V$ at least as small as $\lambda $.
 In view of   \rep{as}, it is clear that to assign the convergence rate of the state transition matrix of
$\bar{A}(I_{mn}-P(I_{mn}-\bar{S}(\tau)))^q$ it is necessary and
sufficient to control the convergence rate
 of the state transition matrix of $A_{V}(\tau)$. This can be accomplished by choosing $q$ sufficiently large.
There are two different ways to do this, each utilizing a different
matrix norm. Both approaches
 will be explained next
using the abbreviated notation  $B(\tau) = V'\bar{S}(\tau)V$; note
that with this simplification, $A_V(\tau) = \tilde{A}B^q(\tau)$
because of \rep{ench}.

\subsection{Weighted Two-Norm}
For each fixed $\tau$ and each appropriately-sized matrix $M$, write
  $\|M\|_{R(\tau)}$ for the matrix norm  induced   by the vector norm
$\|Mx\|_{R(\tau)} \dfb \sqrt{x'R(\tau )x}$. Note that
$\|M\|_{R(\tau)}$ is the largest singular value of
 $R^{\frac{1}{2}}(\tau)MR^{-\frac{1}{2}}(\tau)$.
 Note in addition that
$$ (R^{\frac{1}{2}}(\tau)B(\tau)R^{-\frac{1}{2}}(\tau))'
  (R^{\frac{1}{2}}(\tau)B(\tau) R^{-\frac{1}{2}}(\tau)) < I
 $$
because of \rep{ly}. This shows that the largest singular value of
$R^{\frac{1}{2}}(\tau)B(\tau) R^{-\frac{1}{2}}(\tau)$ is less than
one. Therefore \eq{\|B(\tau)\|_{R(\tau)}<1\label{ino}}
\subsubsection{$\mathbb{N}$ is constant}
In this case both $B(\tau)$  and $R(\tau)$ are constant so it is
sufficient so choose
 choose $q$ so that
$\|\tilde{A}B^q(\tau)\|_{R(\tau)}\leq \lambda $. Since
$\|\cdot\|_{R(\tau )}$ is submultiplicative, this can be done by
choosing $q$ so that
$$\|B(\tau)\|^q_{R(\tau)} \leq \frac{\lambda}{\;\;\;\;\;\|\tilde{A}\|_{R(\tau)}}$$
This can always be accomplished because of \rep{ino}.

\subsubsection{$\mathbb{N}$ changes with time}
In this case it is not possible to use the weighted two-norm
$\|\cdot\|_{R(\tau )}$ because it is time-dependent. A simple fix,
but perhaps not the most efficient one,  would be to use
 the standard two-norm $\|\cdot\|_2$ instead since it does not depend on time.
 Using this approach, the first step would be to
first choose, for each fixed $\tau$,
 an integer $p_1(\tau )
$ large enough so that $\|B^{p_1(\tau)}(\tau )\|<1$.  Such values of
$p_1(\tau)$  must exist because each $B(\tau)$ is
 a discrete-time stability
 matrix or equivalently, a matrix with a spectral radius less than $1$.
  Computing such a value amounts to looking at the
  largest singular value of $B^{p_1(\tau)}(\tau)$ for successively largest values of  $p_1(\tau)$
   until that singular value is less than $1$. Having accomplished this, a number $p$ can easily be computed
   so that $\|B^p(\tau )\|<1\;\forall \tau$ since there are only a finite number of
   distinct strongly connected graphs on $m$ vertices and consequently only a finite number of distinct
matrices   $B(\tau)$ in the set $\scr{B} = \{B(\tau ):\tau\geq 0\}$.
Choosing $p$ to be the maximum of $p_1(\tau)$ with respect to $\tau
$ is thus a finite computation. The next step would be to compute
 an integer $\bar{p}$
 large enough so that each $\|\tilde{A}(B^{p}(\tau))^{\bar{p}}\|_2\leq \lambda$.
 A value of $q$ with the
 required property would then be $q = p\bar{p}$.

\subsection{Mixed Matrix Norm }
There is a different way to choose $q$ which   does
 not make use of either  Lemma \ref{brian} or Proposition \ref{mp}.  The approach exploits the
  ``mixed matrix norm'' introduced in \cite{lineareqn}. To define this norm requires several steps. \;
To begin, let $\|\cdot\|_{\infty}$
 denote   the standard induced
infinity  norm
    and write
 $\R^{mn\times mn}$ for  the vector space of all $m\times m$  block matrices $M = \matt{M_{ij}}$
whose $ij$th entry     is a matrix $M_{ij}\in\R^{n\times n}$. With
$n_i = \dim \scr{V}_i,\;i \in\mathbf{m},$ and $\bar{n} =
n_1+n_2+\cdots n_m$,
 write $\R^{mn\times \bar{n}}$
for the vector space of all $m\times m$  block matrices $M =
\matt{M_{ij}}$ whose $ij$th entry     is a matrix
$M_{ij}\in\R^{n\times n_j}$. Similarly write $\R^{ \bar{n}\times
mn}$ for the vector space of all $m\times m$  block matrices $M =
\matt{M_{ij}}$ whose $ij$th entry     is a matrix
$M_{ij}\in\R^{n_i\times n}$. Finally write
 $\R^{\bar{n}\times \bar{n}}$ for  the vector space of all $m\times m$  block matrices $M = \matt{M_{ij}}$
whose $ij$th entry     is a matrix $M_{ij}\in\R^{n_i\times n_j}$.

Note that $B\in \R^{mn\times mn}$, $\tilde{A}\in \R^{\bar{n}\times
\bar{n}}$, $V\in \R^{mn\times \bar{n}}$, and $V'\in\R^{\bar{n}\times
mn}$.
 For $M$ in any one of these four spaces, the {\em mixed matrix norm } \cite{lineareqn} of $M$, written $\|M\|$, is
\eq{\|M\| = \|\langle M\rangle \|_{\infty}\label{mmn}}
 where $\langle M\rangle $ is the  matrix in $\R^{m\times m}$  whose $ij$th entry is $\|M_{ij}\|_2$.
It is very easy to verify that $\|\cdot\|$ is in fact a norm. It is
even sub-multiplicative whenever matrix multiplication is defined.
Note in addition that $\|V\| = 1$ and $\|V'\| = 1$ because the
columns of each $V_i$ form an  orthonormal set.

Recall that $P = VV'$ is an orthogonal projection matrix. Using
this, the definition of $B(\tau)$ and the fact that $PV = V$, it is
easy to see that for any integer $p>0$
$$B^{p}(\tau) =V'(P\bar{S}(\tau)P)^pV $$
Thus
$$ \|B^p(\tau)\|\leq \|(P\bar{S}(\tau)P)^{p}\|$$
Using this and  the  fact that the graph of $S'$ is strongly
connected,  one can conclude that
$$\|(P\bar{S}(\tau)P)^{p}\|<1,\;\;\;p\geq (m-1)^2$$
This is a direct consequence of  Proposition
 2 of \cite{lineareqn}. Thus
 \eq{\|B^p(\tau)\| <1 ,\;\;\;p\geq (m-1)^2\label{rajit}}

\subsubsection{$\mathbb{N}$ is constant}
In this case  $B(\tau)$  is constant so it is sufficient to choose
$q$ so that
 $\|\tilde{A}B^q(\tau)\|\leq \lambda $.
This can be done by choosing $q = p\bar{p}$  where $p\geq (m-1)^2$
and $\bar{p}$ is such that
 \eq{\|B^p(\tau)\|^{\bar{p}} \leq \frac{\lambda}{\|\tilde{A}\|}\label{rajit2}}
This can always be accomplished because of \rep{rajit}.

\subsubsection{$\mathbb{N}$ changes with time}
Note that \rep{rajit} holds for all $\tau $. Assuming $p$ is chosen
so that  $p\geq (m-1)^2$ it is thus possible to find, for each
$\tau$,  a positive integer $\bar{p}(\tau)$, for which
\eq{\|B^p(\tau)\|^{\bar{p}(\tau)} \leq
\frac{\lambda}{\|\tilde{A}\|}\label{rajit3}}

   Having accomplished this, a number $\bar{p}$ can easily be computed
   so that
\eq{\|B^p(\tau)\|^{\bar{p}} \leq
\frac{\lambda}{\|\tilde{A}\|}\label{rajit3}} holds for all $\tau$,
 since there  there are only a finite number of
   distinct strongly connected graphs on $m$ vertices and consequently only a finite number of distinct
matrices   $B(\tau)$ in the set $\scr{B} $ defined earlier.
 Choosing $\bar{p}$ to be the maximum of
$\bar{p}(\tau)$ with respect to $\tau $ is thus a finite
computation.
 A value of $q$ with the
 required property would then be $q = p\bar{p}$.

\section{Concluding Remarks}

The state estimator developed in this paper relies on an especially
useful observation about distributed observer structure
  first noted in \cite{Kim2016CDC} and subsequently exploited in \cite{trent} and \cite{ACC19}.
  Just how much further this idea
  can be advanced remains to be seen. For sure, the synchronous switching upon which the local
  estimators in this paper depend, can be relaxed by judicious application
  of the mixed matrix norm discussed here. This generalization will be addressed in a future paper.

\bibliographystyle{unsrt}
\bibliography{ACC18,my,steve}

\end{document}